# STUDY OF THE C IV (λλ1548.187 - 1550.772), Si IV (λλ1393.755 - 1402.770) AND O IV (λ1401.156) REGIONS OF THE QSO J021327.25-001446.93

*Ch. Katsavrias, E. Danezis and A. Antoniou*

*National and Kapodistrian University of Athens, Faculty of Physics, Section of Astrophysics, Astronomy and Mechanics, Panepistimiopolis, Zografos, Athens, Greece.*

ABSTRACT

Broad Absorption Line Regions - BALR are composed of a number of successive independent absorbing density layers. Using the GR model, we analyze the UV Si IV (λλ1393.755 - 1402.770), O IV (λ1401.156) and C IV (λλ1548.187 - 1550.772) resonance lines in the spectra of a certain QSO and discuss the results concerning its kinematic properties (rotational, radial and random velocities).

INTRODUCTION

Quasar absorption lines are separated in two major categories concerning their line widths. The narrow absorption lines (NALs) have line widths less than a few hundred km/s and tend to have relatively sharp profiles. The broad absorption lines (BALs) are displaying deep, broad, and smooth absorption troughs extending blueward of the emission-line profile of the absorbing species. In the case of BALs there is no doubt that the absorbing gas is associated with the quasars because it is outflowing at sub relativistic speeds: outflow velocities of order 20,000 to 30,000 km/s are not uncommon, and velocities reaching 60,000 km/s have been detected (Jannuzi et al. 1996; Hamann et al. 1997). The spectrum of a Broad Absorption Line Quasar (BALQSO) is usually interpreted as a combination of a broadband continuum arising from the central engine, the broad emission lines coming from the Broad Emission Line Region (BELR), emerging near the center of the QSO and the broad absorption lines that are superposed, originating in a separate outlying region – Broad Absorption Line Region (BALR) (Lyratzi et al, 2009) .

The spectrum of a BALQSO shows high ionization species as C IV λ1549 and Si IV λ1397 (Hamann et al., 1993; Crenshaw et al., 2003). The purpose of this work is to study the spectrum of a BALQSO regarding the C IV and Si IV lines in order to export important results concerning their kinematic properties.

DATA AND METHOD

The spectrum of the BALQSO – J021327.25-001446.93 that is used in this study has been derived from the SDSS database. Certain facts of this quasar are in detail in Table 1.

| SpecObjID | 114219914419503104 |
|---|---|
| RA-DEC (decimal) | 33.3635527, -0.24636965 |
| Plate | 405 |
| Mjd | 51816 |
| fiberId | 197 |
| Z | 2.3995 |

TABLE 1. Identification of the quasar.

Assuming that the BALR and BELR are composed of a number of successive independent absorbing/emitting density layers of matter (that are originated in a disk wind), which have apparent rotational and radial velocities and where ions have random velocities (Lyratzi et al, 2009), we used the GR model (Danezis et al, 2003 and 2007) to fit the absorption lines of C IV and Si IV. From this analysis we can calculate the values of a group of physical parameters, such as the apparent rotational and radial velocities, the random velocities of the thermal motions of the ions, the Full Width at Half Maximum (FWHM), the optical depth, as well as the absorbed energy and the column density. The respective emission lines are fitted with a Voigt distribution.

The normalized intensity and the fit are shown in figure 1.

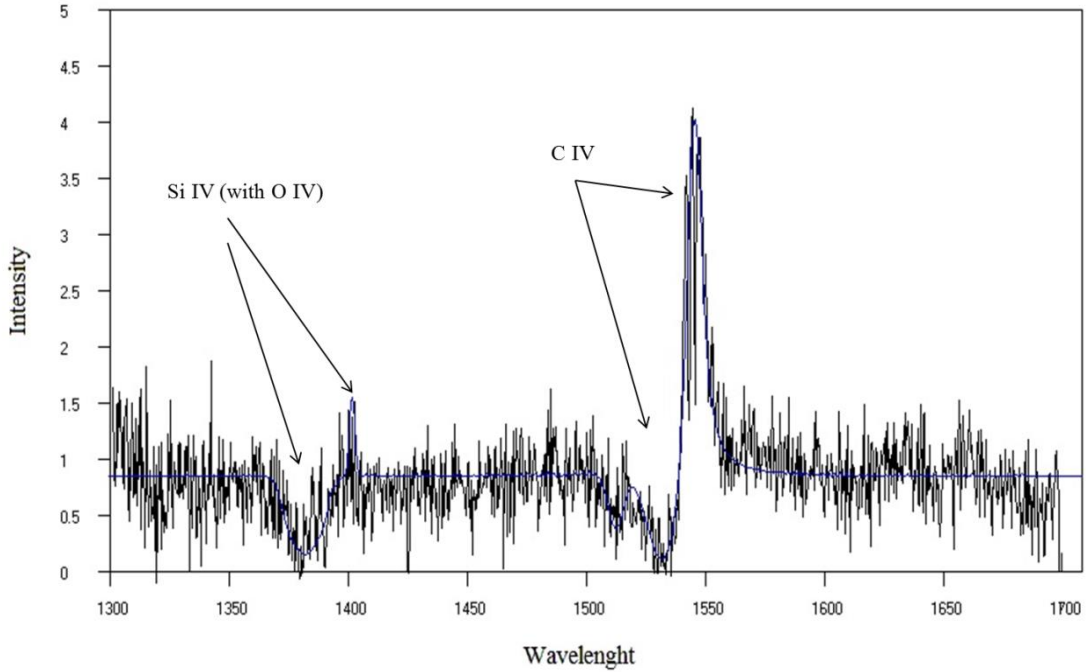

Figure 1. BALRs of the QSO - 114219914419503104. Black line represents the normalized intensity while blue line represents the fit. Emissions are fitted with Voight distribution and absorptions with GR model. There is an absorption line due to existence of O IV which is covered by the Si.

RESULTS AND DISCUSSION

As one can see from Fig. 1 the model is able to fit lines assuming one or more absorbing components. There are two absorption lines and one emission line at the region of C IV. The Si IV region shows an absorption and an emission line which coexist with an O IV absorption line. All the absorption lines are fitted with GR model and the emission lines with Voight (C IV) and Gauss (Si IV) distribution.

The results of the best fit are presented in Table 2 where the random and rotational velocities as well as radial (outflow) velocity, the Full Width at Half Maximum (FWHM), the optical depth, as well as the absorbed energy and the column density are given.

For the C IV absorption lines we found random velocities between 637 and 1346 Km/s which agrees with previous results for random velocities by Lyratzi et al, 2009. We also notice the comparable random velocities of the Si IV emission line and the O IV absorption line. The rotation velocity of Si IV absorption lines are 400 Km/sec while the respective lines of C IV have rotation velocities of 50 Km/sec even though the FWHM is larger.

| Line | Fit | Vrand (Km/sec) | Vrot (Km/sec) | Vrad (Km/sec) | FWHM | CD ($10^{10}$/cm$^2$) | Eabs (eV) | V MIX | Opt. Depth | S |
|---|---|---|---|---|---|---|---|---|---|---|
| C IV (1548.187) | Voigt | - | 3 | -832.66 | 6.649 | 13.721 | 8.0083 | 5.7 | 1.35 | 2.8 |
| | RG | 1346.101 | 50 | -3291.902 | 15.796 | -4.787 | | 5.9 | 0.75 | - |
| | RG | 638.8275 | 1 | -7164.721 | 7.074 | -1.353 | | 2.8 | 0.40 | - |
| C IV (1550.772) | Voigt | - | 3 | -831.27 | 6.451 | 12.683 | 7.995 | 5.7 | 1.215 | 2.8 |
| | RG | 1343.857 | 50 | -3286.415 | 15.607 | -4.416 | | 5.9 | 0.675 | - |
| | RG | 637.7626 | 1 | -7152.784 | 7.025 | -1.236 | | 2.8 | 0.360 | - |

| Line | Fit | Vrand (Km/sec) | Vrot (Km/sec) | Vrad (Km/sec) | FWHM | CD ($10^{10}$/cm$^2$) | Eabs (eV) | V MIX | Opt. Depth | S |
|---|---|---|---|---|---|---|---|---|---|---|
| Si IV (1393.755) | Gauss | 253.4324 | 1 | -258.1167 | 2.947 | 0.56905 | 8.8957 | - | 1.40 | 1 |
| | RG | 937.7008 | 400 | -3441.556 | 10.853 | -1.6234 | | 3.7 | 1.15 | - |
| Si IV (1402.770) | Gauss | 253.8039 | 1 | -256.4579 | 2.862 | 0.5157 | 8.8385 | - | 1.19 | 1 |
| | RG | 931.674 | 400 | -3419.439 | 10.583 | -1.46 | | 3.7 | 0.978 | - |

| Line | Fit | Vrand (Km/sec) | Vrot (Km/sec) | Vrad (Km/sec) | FWHM | CD ($10^{10}$/cm$^2$) | Eabs (eV) | V MIX | Opt. Depth | S |
|---|---|---|---|---|---|---|---|---|---|---|
| O IV (1401.156) | RG | 252.094 | 1 | -1840.066 | 3.026 | -0.92533 | 8.8487 | 1 | 1.6 | - |

TABLE 2. Fitting parameters of C IV, Si IV and O IV lines.